\newif\ifpreprint

\preprintfalse 

\ifpreprint
\documentclass[journal=jctcce,manuscript=letter]{achemso}
\else
\documentclass[journal=jctcce,manuscript=letter,layout=twocolumn]{achemso}
\fi

\usepackage[T1]{fontenc} 

\usepackage{amsmath}
\usepackage{newtxtext,newtxmath}

\usepackage{graphicx}
\usepackage{dcolumn}
\usepackage{braket}
\usepackage{multirow}
\usepackage{threeparttable}
\usepackage{xspace}
\usepackage{verbatim}
\usepackage[version=4]{mhchem} 
\usepackage{comment}
\usepackage{color,soul}
\usepackage{siunitx}
\usepackage{physics}

\usepackage{mathtools}
\usepackage[dvipsnames]{xcolor}
\usepackage{xspace}
\usepackage{ifthen}

\usepackage{qcircuit}

\usepackage{graphicx,longtable,dcolumn,mhchem}
\usepackage{rotating,color}
\usepackage{lscape}
\usepackage{amsmath}
\usepackage{dsfont}
\usepackage{soul}
\usepackage{physics}
\newcolumntype{d}{D{.}{.}{-1}}

\newcommand{\ie}{\textit{i.e}}

\newcommand{\ra}{\rightarrow}
\newcommand{\pis}{\pi^\star}
\newcommand{\npi}{n\ra\pis}
\newcommand{\ppi}{\pi\ra\pis}
\newcommand{\pts}{\pi\ra{3s}}
\newcommand{\nts}{n\ra{3s}}
\newcommand{\ptp}{\pi\ra{3p}}
\newcommand{\ntp}{n\ra{3p}}

\newcommand{\Pop}{6-31+G(d)}
\newcommand{\AVDZ}{{aug}-cc-pVDZ}
\newcommand{\AVTZ}{{aug}-cc-pVTZ}

\usepackage[normalem]{ulem}

\usepackage[colorlinks = true,
            linkcolor = blue,
            urlcolor  = blue,
            citecolor = blue,
            anchorcolor = blue]{hyperref}
\definecolor{goodorange}{RGB}{225,125,0}
\definecolor{goodgreen}{RGB}{5,130,5}
\definecolor{goodred}{RGB}{220,50,25}
\definecolor{goodblue}{RGB}{30,144,255}

\newcommand{\note}[2]{
\ifthenelse{\equal{#1}{F}}{
\colorbox{goodorange}{\textcolor{white}{\footnotesize \fontfamily{phv}\selectfont #1}}
    \textcolor{goodorange}{{\footnotesize \fontfamily{phv}\selectfont #2}}\xspace
}{}
\ifthenelse{\equal{#1}{R}}{
\colorbox{goodred}{\textcolor{white}{\footnotesize \fontfamily{phv}\selectfont #1}}
    \textcolor{goodred}{{\footnotesize \fontfamily{phv}\selectfont #2}}\xspace
}{}
\ifthenelse{\equal{#1}{N}}{
\colorbox{goodgreen}{\textcolor{white}{\footnotesize \fontfamily{phv}\selectfont #1}}
    \textcolor{goodgreen}{{\footnotesize \fontfamily{phv}\selectfont #2}}\xspace
}{}
\ifthenelse{\equal{#1}{M}}{
\colorbox{goodblue}{\textcolor{white}{\footnotesize \fontfamily{phv}\selectfont #1}}
    \textcolor{goodblue}{{\footnotesize \fontfamily{phv}\selectfont #2}}\xspace
}{}
}

\usepackage{titlesec}

\usepackage[fontsize=11pt]{scrextend}
\titleformat{\section}
{\normalfont\sffamily\bfseries\color{Blue}}
{\thesection.}{0.25em}{\uppercase}

\titleformat{\subsection}
{\normalfont\sffamily\bfseries}
{\thesubsection}{0.25em}{}

\titleformat{\subsubsection}
{\normalfont\sffamily}
{\thesubsubsection}{0.25em}{}

\titleformat{\suppinfo}
{\normalfont\sffamily\bfseries}
{\thesubsection}{0.25em}{}

\titlespacing*{\section}{0pt}{0.5\baselineskip}{0.01\baselineskip}
\titlespacing*{\subsection}{0pt}{0.125\baselineskip}{0.01\baselineskip}
\titlespacing*{\subsubsection}{0pt}{0.125\baselineskip}{0.01\baselineskip}

\newcommand{\LCPQ}{Laboratoire de Chimie et Physique Quantiques, Universit\'e de Toulouse, CNRS, UPS, France}
\newcommand{\SMU}{Department of Chemistry, Southern Methodist University, Dallas, Texas 75275, USA}
\newcommand{\UOP}{Dipartimento di Chimica e Chimica Industriale, University of Pisa, Via Moruzzi 3, 56124 Pisa, Italy}
\newcommand{\CEISAM}{Nantes Universit\'e, CNRS,  CEISAM UMR 6230, F-44000 Nantes, France}

\author{Pierre-Fran{\c c}ois Loos}
	\email{loos@irsamc.ups-tlse.fr}
	\affiliation[LCPQ, Toulouse]{\LCPQ}
\author{Filippo Lipparini}
	\affiliation[UP, Pisa]{\UOP}
\author{Devin A.~Matthews}
	\affiliation[SMU, Dallas]{\SMU}
\author{Aymeric Blondel}
	\affiliation[UN, Nantes]{\CEISAM}    
\author{Denis Jacquemin}
	\email{Denis.Jacquemin@univ-nantes.fr}
	\affiliation[UN, Nantes]{\CEISAM}

\let\oldmaketitle\maketitle
\let\maketitle\relax
     \title{A Mountaineering Strategy to Excited States: Revising Reference Values with EOM-CC4}
\date{\today}

\begin{tocentry}
	\centering
	\includegraphics[width=.65\textwidth]{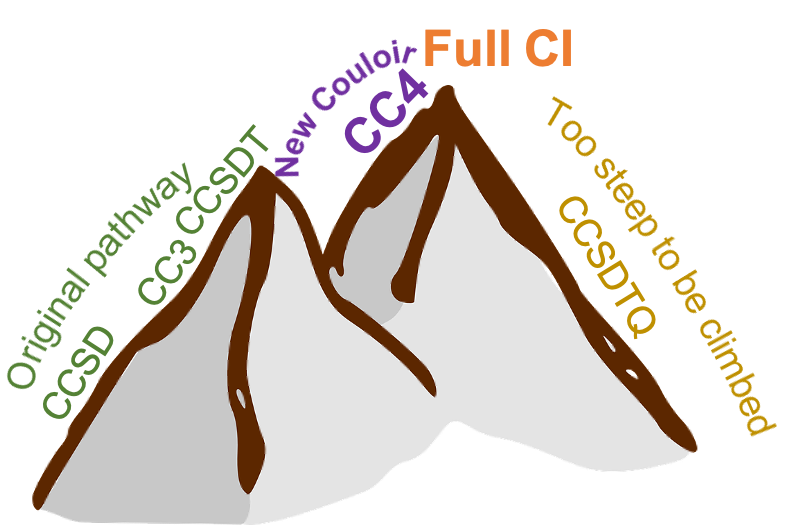}
\end{tocentry}

\begin{document}

\ifpreprint
\else
\twocolumn[
\begin{@twocolumnfalse}
\fi
\oldmaketitle

\begin{abstract}
In the framework of the computational determination of highly-accurate vertical excitation energies in small organic compounds, we explore the possibilities offered by the equation-of-motion formalism 
relying on the approximate fourth-order coupled-cluster (CC) method, CC4.  We demonstrate, using an extended set of more than 200 reference values based on CC including up to quadruples excitations 
(CCSDTQ), that CC4 is an excellent approximation to CCSDTQ for excited states with a dominant contribution from single excitations with an average deviation as small as $0.003$ eV. We next assess 
the accuracy of several additive basis set correction schemes, in which vertical excitation energies obtained with a compact basis set and a high-order CC method are corrected with lower-order CC calculations 
performed in larger basis sets. Such strategies are found to be overall very beneficial, though their accuracy depend significantly on the actual scheme. Finally, CC4 is employed to improve several theoretical 
best estimates of the QUEST database for molecules containing between four and six (non-hydrogen) atoms, for which previous estimates were computed  at the CCSDT level.
\end{abstract}

\ifpreprint
\else
\end{@twocolumnfalse}
]
\fi

\ifpreprint
\else
\small
\fi

\noindent

\section{Introduction}
\label{sec:Intro}

Defining sets of high-quality reference values that can be employed to reliably assess the \emph{pros} and \emph{cons} of lower-cost theoretical methods is a very popular and useful 
research line in quantum chemistry. \cite{Roo96,Pie02,Dre05,Kry06,Sne12,Gon12,Lau13,Ada13a,Gho18,Bla20,Loo20a} Although experimental values may constitute natural references 
for some properties (e.g., thermodynamical and kinetic data), \cite{Pop89,Cur91,Cur97,Cur98,Cur07} it is often welcome to rely on state-of-the-art electronic structure methods to produce 
reference values for other properties. Of course, this latter approach is intrinsically limited by the computational cost of these high-accuracy models. However, it has the undeniable 
advantage to allow well-grounded comparisons within a unique, well-defined set of parameters. \cite{Taj04,Bom06,Har08,Set13} Indeed, one can perform comparisons with exactly the 
same geometries, basis set, solvent model, etc.  Purely theoretical reference values are especially useful for electronic excited states (ESs), as the most straightforwardly accessible 
theoretical values, namely, vertical transition energies (VTEs), are not directly measurable experimentally. \cite{Loo20c} This explains why the determination of accurate VTEs has been 
an active and productive avenue of research during the past three decades, with, in particular, valuable works from Roos' \cite{Ser93,Ser93b,Ser95} and Thiel's  \cite{Sch08,Sil10b,Sil10c} groups.

Since 2018, our groups have made several contributions in this field, \cite{Loo18a,Loo19c,Loo20a,Loo20d,Chr21,Loo21a} and it eventually led to the creation of the QUEST database [see \url{https://lcpq.github.io/QUESTDB_website}], 
that contains a large panel of reference VTEs for molecules containing from 1 to 10 non-hydrogen atoms. \cite{Ver21} At the present stage, the QUEST database includes more than 500 
theoretical best  estimates (TBEs) for diverse ESs  (singlet, doublet, and triplet; valence and Rydberg; charge transfer, singly- and doubly-excited states) that have been established with the {\AVTZ} 
basis set. Typically, the TBEs contained in the QUEST database are produced using VTEs computed with the selected configuration interaction (SCI) algorithm named \textit{``Configuration 
Interaction using a Perturbative Selection made Iteratively''} (CIPSI) \cite{Hur73,Gin13,Gin15,Gar17b,Gar18,Gar19} to obtain near full configuration interaction (FCI) quality VTEs for systems 
containing from 1 to 3 non-hydrogen atoms, coupled cluster (CC) with singles, doubles, triples and, quadruples (CCSDTQ) \cite{Kuc91} for molecules encompassing 4 non-hydrogen atoms, and 
CC with singles, doubles, and triples (CCSDT) \cite{Nog87,Scu88,Kuc01,Kow01,Kow01b} for larger derivatives. Popular basis set correction schemes have often been applied.
For example, the CCSDTQ VTEs computed with a  double-$\zeta$ basis set were corrected thanks to CCSDT values obtained with a triple-$\zeta$ basis set. Most TBEs included in the QUEST database
were estimated to be chemically-accurate (corresponding to 1 kcal.mol$^{-1}$ or $0.04$ eV error), with a typical error bar of $\pm$0.03 eV.  The reference values included 
in QUEST have been used by various groups, for example, to
(i) assess the relative accuracies of third-order \cite{Loo20b}, multireference, \cite{Sar22} and other emerging \cite{Gin19,Oti20,Loo20f,Kos21,Gou21,Gou22} methods, 
(ii) quantify the accuracy of local hybrids for triplet ESs,  \cite{Gro21} 
(iii) determine the relative performance of several hybrid \cite{Lia22} or double hybrid \cite{Cas19,Mes21,Cas21b,Mes21b} functionals, and 
(iv) evaluate the potential of orbital-optimized  density-functional theory for double excitations. \cite{Hai20b,Hai21}

Despite burning an unreasonable number of CPU hours during the past four years, we could hardly ``do better'' than what is described above, as we rapidly hit the computational wall of both 
high-order CC schemes and/or large CIPSI calculations. For instance, CCSDTQ formally scales as $\order*{N^{10}}$ (where $N$ is the number of basis functions) and 
determining VTEs in molecules like furan or thiophene is nearly impossible even in a double-$\zeta$ basis set.  In an effort to go one step further we explored very recently, and for the first time, 
the performance of the approximate fourth-order CC model, CC4, \cite{Kal05} in the context of ESs. \cite{Loo21b} From a theoretical point of view, CC4 can be viewed as an approximation of 
CCSDTQ that still includes iterative quadruples, but neglects the calculation of the most expensive components (and avoids the storage of the higher-excitation amplitudes), allowing to reduce 
the formal scaling by one order of magnitude to $\order*{N^{9}}$. \cite{Kal05} In our first investigation, \cite{Loo21b} we considered very small systems (\ce{BH}, \ce{BF}, \ce{CO}, \ce{HCl}, 
\ce{H2O}, \ce{H2S}, \ce{N2}, and \ce{NH3}) for which well-converged CIPSI calculations were achievable for 25 ESs. This preliminary study indicates that CC4  indeed provides highly competitive 
VTEs, as it allows to significantly reduce the CCSDT error as compared to FCI, with final deviations --- for ESs with a dominant single excitation character --- only slightly larger than their 
CCSDTQ counterparts.

In the present contribution, which is only the second work presenting CC4 calculations for ESs, we build on our previous study,  \cite{Loo21b} and we aim at 
(i) evaluating the performance of CC4 for a more significant set of molecules and a larger variety of ESs, 
(ii) assessing the accuracy of additive basis set correction procedures based on CC4, and 
(iii) providing improved TBEs for many ESs included in the QUEST database.  

\section{Computational details}
\label{sec:Comp}

\begin{figure}[htp]
  \includegraphics[scale=0.62,viewport=1.2cm 4.1cm 17.3cm 27.2cm,clip]{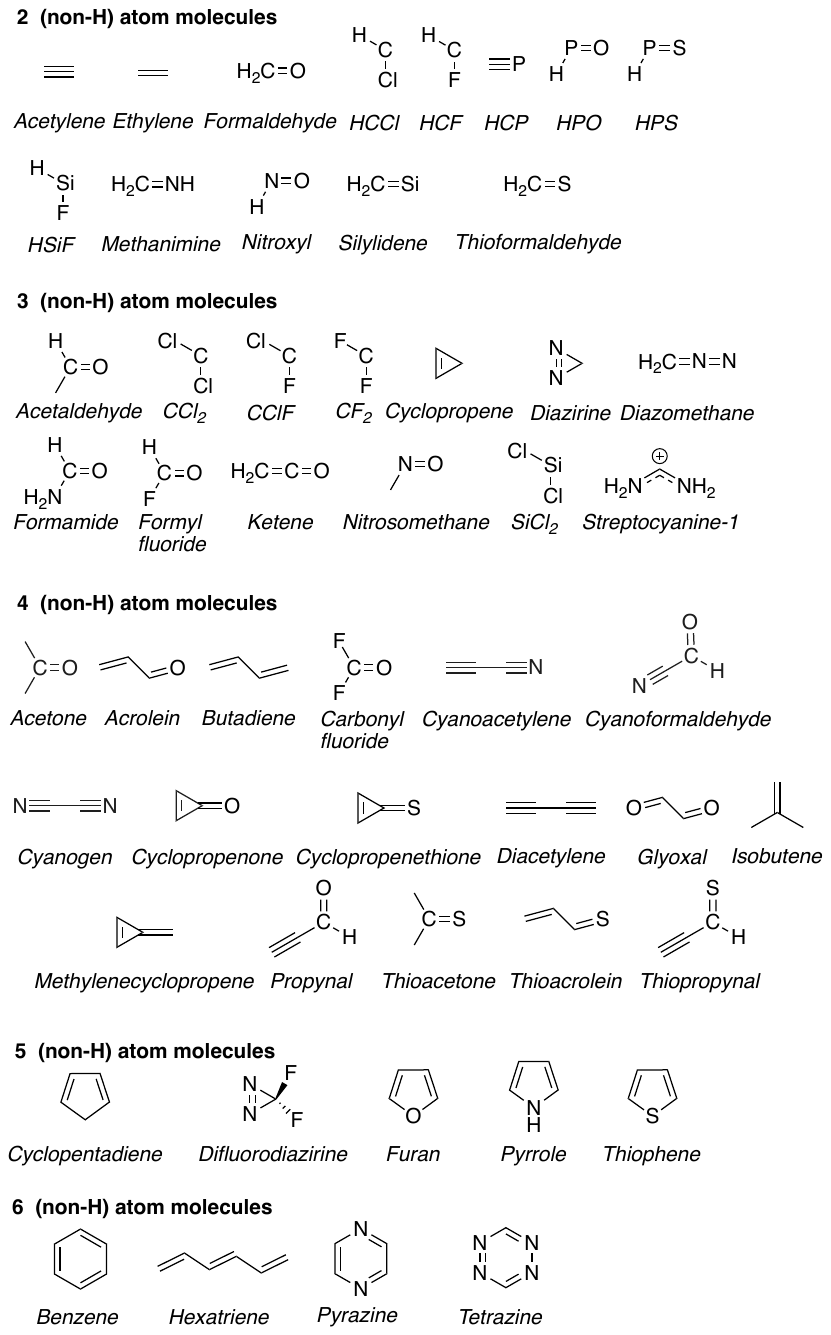}
  \caption{Representation of the systems investigated in the present study.}
  \label{Fig-1}
\end{figure}

All calculations presented here rely on the frozen-core approximation and the high-quality ground-state geometries extracted from the QUEST database. \cite{Ver21}  For the sake of reproducibility, cartesian coordinates 
for all systems displayed in Figure \ref{Fig-1} are reproduced in the supporting information (SI). Note that, below, we do not specify the equation-of-motion  (EOM) prefix for the CC calculations, although all ES calculations 
are performed using this formalism.

Most of our CC calculations have been performed with CFOUR,  \cite{cfour,Mat20}  that provides an efficient implementation of high-order CC methods up to quadruples. 
\cite{Mat15a,Mat19} For all considered systems, we performed, when technically achievable, CC2, \cite{Chr95,Hat00} CCSD, \cite{Pur82,Scu87,Koc90b,Sta93,Sta93b} CC3, 
\cite{Chr95b,Koc95,Koc97}  CCSDT, \cite{Nog87,Scu88,Kuc01,Kow01,Kow01b} CC4, \cite{Kal04,Kal05} and CCSDTQ \cite{Kuc91,Kal04} calculations using three Gaussian 
basis sets: {\Pop}, {\AVDZ}, and {\AVTZ}. For the sake of conciseness, these basis sets are sometimes dubbed as ``Pop'', ``AVDZ'', and ``AVTZ'' in the following. {These three 
bases were selected because they contain both polarization and diffuse, and are widely available, AVTZ being typically viewed as providing results close from the basis set limit 
for transition energies, but for very diffuse Rydberg states.}

{The identification of all states follows the QUEST database.}\cite{Ver21}  {The separation between valence and Rydberg transitions was performed by examining the
dominant unoccupied orbital contribution in the CCSD and CC3 results, which typically gives a clear trend, but for a few high-lying ESs for which mixing can appear. The 
double-excitation character of all transitions was probed by using} $\%T_1$ {the percentage of single excitations as given by CC3. Note that genuine or pure double ESs 
have a negligible}  $\%T_1$ {and are unseen by CC2 and CCSD approaches.}

In the statistical analysis presented below, we report the usual statistical indicators: the mean signed error (MSE), the mean absolute error (MAE), as well as largest positive and 
negative deviations [Max(+) and Max($-$), respectively].

CCSDTQP/{\Pop} calculations have been performed with MRCC \cite{mrcc,Kal20} for the molecules encompassing two non-hydrogen atoms (see  Figure \ref{Fig-1}). These values 
are reported in Table S1 of the SI. Considering all 32 VTEs and taking these CCSDTQP values as references, we obtain a MAE of \SI{0.003}{\eV} for the CCSDTQ data determined 
with the same basis set. Unsurprisingly, the largest error, \SI{0.023}{\eV}, comes from the pure $(n,n) \ra (\pi^\star,\pi^\star)$ doubly-excited state of nitroxyl for which the CC 
expansion obviously converges slower.  Removing this pathological case yields a MAE of \SI{0.002}{\eV} with all errors below \SI{0.010}{\eV}, clearly confirming the quality of the 
CCSDTQ estimates.

\begin{table*}[htp]
\caption{\footnotesize Statistical analysis obtained using various correction schemes. MSE, MAE and maximal deviations (in \si{\eV}) are obtained with respect to the reference method given in the leftmost column.
The number of reference ESs considered at the CCSDTQ/{\AVTZ}, CC4/{\AVTZ}, and CCSDT/{\AVTZ} levels is 31, 59, and 118, respectively. In the rightmost column, 
we provide the number of states for which the correction provides smaller (or equal) absolute errors as compared to the non-corrected scheme (rows labeled as ``no correction'').}
\label{Table-1}
\footnotesize
\begin{tabular}{p{3.5cm}p{3.5cm}p{5cm}ccccc}
ine
Approximating\ldots	&using\ldots		& and correcting with\ldots		&MSE	&MAE	&Max(+)	&Max($-$)	&Useful?\\
ine															
CCSDTQ/{\AVTZ}	&CC4/{\AVTZ}	&	no correction					&0.002	&0.003	&0.012	&-0.003	&	\\
				&			&	+ [CCSDTQ $-$ CC4]/{\AVDZ}		&0.000	&0.001	&0.003	&-0.003	&22/31 	\\
				&			&	+ [CCSDTQ $-$ CC4]/{\Pop}		&0.001	&0.002	&0.006	&-0.004	&19/31	\\ 
				&CCSDT/{\AVTZ}	&	no correction				&-0.001	&0.017	&0.107	&-0.049	&	\\
				&			&	+ [CCSDTQ $-$ CCCST]/{\AVDZ}	&0.003	&0.004	&0.022	&-0.004	&27/31 	\\
				&			&	+ [CCSDTQ $-$ CCSDT]/{\Pop}	&0.001	&0.003	&0.030	&-0.005	&25/31	\\ 
				&CC3/{\AVTZ}	&	no correction					&0.006	&0.017	&0.124	&-0.034	&	\\
				&			&	+ [CCSDTQ $-$ CC3]/{\AVDZ}		&0.002	&0.004	&0.019	&-0.003	&28/31 	\\
				&			&	+ [CCSDTQ $-$ CC3]/{\Pop}		&0.001	&0.004	&0.020	&-0.008	&26/31	\\ 
				&CCSD/{\AVTZ}	&	no correction				&0.071	&0.071	&0.273	&-0.005	&	\\
				&			&	+ [CCSDTQ $-$ CCSD]/{\AVDZ}	&0.022	&0.024	&0.066	&-0.011	&26/31 	\\
				&			&	+ [CCSDTQ $-$ CCSD]/{\Pop}		&0.010	&0.032	&0.061	&-0.042	&25/31	\\ 
				&CC2/{\AVTZ}	&	no correction					&-0.008	&0.215	&0.348	&-0.695	&	\\
				&			&	+ [CCSDTQ $-$ CC2]/{\AVDZ}		&0.032	&0.032	&0.196	&-0.001	&29/31 	\\
				&			&	+ [CCSDTQ $-$ CC2]/{\Pop}		&-0.005	&0.051	&0.108	&-0.353	&29/31	\\ 
CC4/{\AVTZ}		&CCSDT/{\AVTZ}	&	no correction				&0.004	&0.016	&0.095	&-0.055	&	\\
				&			&	+ [CC4 $-$ CCSDT]/{\AVDZ}		&0.004	&0.004	&0.022	&-0.005	&52/59 	\\
				&			&	+ [CC4 $-$ CCSDT]/{\Pop}		&0.002	&0.004	&0.028	&-0.009	&55/59	\\ 
				&CC3/{\AVTZ}	&	no correction					&0.006	&0.016	&0.112	&-0.043	&	\\
				&			&	+ [CC4 $-$ CC3]/{\AVDZ}			&0.002	&0.003	&0.019	&-0.004	&52/59 	\\
				&			&	+ [CC4 $-$ CC3]/{\Pop}			&0.000	&0.004	&0.018	&-0.010	&52/59	\\ 
				&CCSD/{\AVTZ}	&	no correction				&0.083	&0.084	&0.261	&-0.009	&	\\
				&			&	+ [CC4 $-$ CCSD]/{\AVDZ}		&0.018	&0.019	&0.063	&-0.009	&54/59 	\\
				&			&	+ [CC4 $-$ CCSD]/{\Pop}			&0.008	&0.023	&0.081	&-0.041	&51/59 	\\
				&CC2/{\AVTZ}	&	no correction					&0.053	&0.193	&0.427	&-0.701	&	\\
				&			&	+ [CC4 $-$ CC2]/{\AVDZ}			&0.027	&0.029	&0.196	&-0.016	&56/59 	\\
				&			&	+ [CC4 $-$ CC2]/{\Pop}			&0.000	&0.049	&0.183	&-0.352	&54/59	\\ 
CCSDT/{\AVTZ}	&CC3/{\AVTZ}	&	no correction					&0.004	&0.015	&0.073	&-0.046	&	\\
				&			&	+ [CCSDT $-$ CC3]/{\AVDZ}		&-0.002	&0.003	&0.007	&-0.016	&109/118	\\
				&			&	+ [CCSDT $-$ CC3]/{\Pop}		&-0.002	&0.005	&0.011	&-0.036	&102/118	\\
				&CCSD/{\AVTZ}	&	no correction				&0.114	&0.114	&0.539	&0.010	&	\\
				&			&	+ [CCSDT $-$ CCSD]/{\AVDZ}		&0.022	&0.025	&0.073	&-0.015	&115/118	\\
				&			&	+ [CCSDT $-$ CCSD]/{\Pop}		&0.012	&0.027	&0.071	&-0.044	&112/118	\\
				&CC2/{\AVTZ}	&	no correction					&-0.007	&0.191	&0.495	&-0.675	&	\\
				&			&	+ [CCSDT $-$ CC2]/{\AVDZ}		&0.020	&0.023	&0.174	&-0.020	&115/118	\\
				&			&	+ [CCSDT $-$ CC2]/{\Pop}		&-0.019	&0.047	&0.167	&-0.355	&106/118	\\
ine
\end{tabular}
\end{table*}
\section{Methodological aspects}
\label{sec:Res1}

\subsection{How does CC4 compare to CCSDTQ?}

Given that CC4 is an approximation of CCSDTQ, and that the accuracy of the latter is recognized as exceptional for ESs with a dominant contribution from singly-excited determinants (see above), 
it seems natural to assess the performance  of CC4 with respect to CCSDTQ. Considering the values listed in the SI (with the three basis sets and all compounds), we have 220 CCSDTQ reference 
values at hand. Using these reference values, one obtains a MSE of \SI{0.002}{\eV} and a MAE as small as \SI{0.003}{\eV} for CC4. As one could have foreseen, \cite{Loo21b} the largest deviations 
between CC4 and CCSDTQ originate from the genuine double excitations in nitroxyl, nitrosomethane, and glyoxal for which CC-based methods would likely not be chosen in practical applications. 
Removing these three states, but keeping all other cases including the $2^1A_g$ dark state of butadiene, leads to a MSE of \SI{0.001}{\eV} and a MAE of \SI{0.003}{\eV} for CC4. Interestingly, in 
our earlier work devoted to very small compounds, we obtained the same MAE of \SI{0.003}{\eV} for CC4 (against CCSDTQP). \cite{Loo21b} Clearly, one can state that CC4 is a very good 
approximation of CCSDTQ for the ESs that are not dominated by doubly-excited determinants.

\subsection{Is CC4 significantly more accurate than CCSDT and CC3?}

The next natural question is to determine if CC4 is worth its cost. Indeed, CC4 formally scales as $\order*{N^9}$ which is more costly than both CCSDT and CC3 [which scale as $\order*{N^8}$ 
and $\order*{N^7}$, respectively].  Using again the available CCSDTQ values as references, we obtain MSE and MAE of \SI{0.000} and \SI{0.015}{\eV} for CCSDT, and \SI{0.006}{} and \SI{0.018}{\eV} 
for CC3, again discarding from the set the true double excitations of nitroxyl, nitrosomethane, and glyoxal. It is no surprise that, on the one hand,  both CC3 and CCSDT deliver results that 
would be rated as accurate enough for many applications, and, on the other hand,  CC3 is an excellent approximation of CCSDT. \cite{Kan14,Taj19,Ver21} The most valuable observation for 
our purposes is that CC4 indeed lowers the CCSDT and CC3 deviations by a factor of five, which we consider a very significant improvement when ones aims at producing highly-accurate 
reference values.

\subsection{Are CC4 basis set effects transferable?}

When defining values for benchmarking purposes, it is obvious from the discussion above that one can safely employ CC4/{\AVTZ} as reference when neither well-converged CIPSI/{\AVTZ} 
nor CCSDTQ/{\AVTZ}  calculations are technically achievable. As detailed below, this is the case, for example, for some of the molecules encompassing four non-hydrogen atoms.  However, 
as customary in the field, \cite{Kal04,Bal06,Kam06b,Pea12,Wat12,Fel14,Gui18b,Fra19,Cas19,Chr21,Loo21a} one can employ a double-$\zeta$ VTE computed with a CC model 
including quadruples in order to correct a triple-$\zeta$ VTE obtained with a quadruple-free CC method. 

To test such strategies, we report in Table \ref{Table-1} the statistical analysis obtained with various correction schemes. We consider as reference the actual results obtained with 
the target method (leftmost column of Table  \ref{Table-1}). These reference values are available in Tables S3, S6, and S9 of the SI. 
In the rightmost column, we provide the number of states for which the correction yields smaller (or equal) absolute errors as compared to the non-corrected scheme (rows labeled as ``no correction'').

Globally, the data listed in Table \ref{Table-1} are quite appealing and one notices three general trends. First, adding corrections for higher-order excitations systematically decrease the MAE and improves the (vast) majority of
the estimates. For example, while CC2/{\AVTZ} returns a MAE of \SI{0.215}{\eV} as compared to CCSDTQ/{\AVTZ}, correcting them with the differences 
between the VTEs computed with these two models but with the much smaller {\Pop} basis set allows reducing the error to \SI{0.051}{\eV} 
and improves 29 of the 31 VTEs.  Second, the more refined the starting point, the more accurate the final estimate. For example, using the same approach as above but starting
 from CC3/{\AVTZ} rather than CC2/{\AVTZ} would cut down the MAE from \SI{0.051}{\eV} to  \SI{0.004}{\eV}. Third, performing the correction with {\AVDZ} typically yields smaller 
 statistical deviations than with {\Pop}, though the difference between the two approaches is small.

Given that CC4 is such a stunning approximation of CCSDTQ for ESs with a single excitation character, it is particularly relevant to investigate CC4/{\AVTZ} estimates based 
on CCSDT and CC3, i.e.,
\begin{subequations}
\begin{align}
\label{eq2}
	\Delta \tilde{E}_{\text{AVTZ}}^{\text{CC4}} 
	& \simeq \Delta E_{\text{AVTZ}}^{\text{CCSDT}} + \qty[ \Delta E_{\text{AVDZ}}^{\text{CC4}} - \Delta E_{\text{AVDZ}}^{\text{CCSDT}} ], 
\\
\label{eq3}
	\Delta \tilde{E}_{\text{AVTZ}}^{\text{CC4}} 
	& \simeq \Delta E_{\text{AVTZ}}^{\text{CCSDT}} + \qty[ \Delta E_{\text{Pop}}^{\text{CC4}} - \Delta E_{\text{Pop}}^{\text{CCSDT}} ], 
\\
\label{eq4}
	\Delta \tilde{E}_{\text{AVTZ}}^{\text{CC4}} 
	& \simeq \Delta E_{\text{AVTZ}}^{\text{CC3}} + \qty[ \Delta E_{\text{AVDZ}}^{\text{CC4}} - \Delta E_{\text{AVDZ}}^{\text{CC3}} ], 
\\
\label{eq5}
	\Delta \tilde{E}_{\text{AVTZ}}^{\text{CC4}} 
	& \simeq \Delta E_{\text{AVTZ}}^{\text{CC3}} + \qty[ \Delta E_{\text{Pop}}^{\text{CC4}} - \Delta E_{\text{Pop}}^{\text{CC3}} ].
\end{align}
\end{subequations}
Looking at Table \ref{Table-1}, one notices that all four approaches deliver very comparable error patterns, with MAE not larger than \SI{0.004}{\eV}, and global improvement over the 
uncorrected CCSDT/{\AVTZ} and CC3/{\AVTZ} data, the errors being cut down by a factor of three. Of course, as can be deduced from the last column, not all values are more accurate 
when adding corrections obtained with smaller basis sets, but improvements are observed for the vast majority of the cases. 

Additionally, although this is not the focus of the present work, we stress that the following expressions
\begin{subequations}
\begin{align}
	\label{eq6}
	\Delta \tilde{E}_{\text{AVTZ}}^{\text{CCSDT}} 
	& \simeq \Delta E_{\text{AVTZ}}^{\text{CC3}} + \qty[ \Delta E_{\text{AVDZ}}^{\text{CCSDT}} - \Delta E_{\text{AVDZ}}^{\text{CC3}} ], 
	\\
	\Delta \tilde{E}_{\text{AVTZ}}^{\text{CCSDT}} 
	& \simeq \Delta E_{\text{AVTZ}}^{\text{CC3}} + \qty[ \Delta E_{\text{Pop}}^{\text{CCSDT}} - \Delta E_{\text{Pop}}^{\text{CC3}} ], 
\end{align}
\end{subequations}
are extremely effective with MAE of \SI{0.003}{} and \SI{0.005}{\eV} and small maximal deviations as compared to the true CCSDT/{\AVTZ} transition energies. The fact that 
these two approximations are very accurate, is certainly valuable, as both CC3/{\AVTZ} and CCSDT/{\Pop} calculations are technically achievable on systems with up to approximately 
10--12 non-hydrogen atoms, allowing to significantly expand the number of TBEs based on CCSDT/{\AVTZ} that are included in the QUEST database for larger systems.

The take-home message of this Section is that, while these correction schemes (which are widespread in the CC community) are generally valuable to produce accurate VTEs, their 
overall accuracy depends ultimately on the quality of the starting point.

\section{Refined theoretical best estimates}
\label{sec:Res2}

Given the above observations, it is worth reexamining our previous TBEs with the additional accuracy provided by CC4.

\subsection{Linear systems}

The QUEST database includes three linear systems encompassing four non-hydrogen atoms: cyanoacetylene, cyanogen, and diacetylene. \cite{Loo20a,Ver21} For these three systems CC4/{\AVTZ} 
calculations could be performed, whereas CCSDTQ/{\AVTZ} calculations remain beyond reach (see Table S9 in the SI for raw data). We have, therefore, computed new TBEs on the basis of these 
CC4 data, applying the very trustworthy basis set correction (see top lines of Table \ref{Table-1})
\begin{equation}
\label{eq8}
	\Delta {E}_{\text{AVTZ}}^{\text{TBE}} 
	= \Delta E_{\text{AVTZ}}^{\text{CC4}} 
	+ \qty[ \Delta E_{\text{AVDZ}}^{\text{CCSDTQ}} - \Delta E_{\text{AVDZ}}^{\text{CC4}} ]. 
\end{equation}
The results for seven ESs of these three linear molecules are listed in Table \ref{Table-2}. It can be seen that the variations are negligible, with three TBEs unchanged, 
and four decreasing by \SI{0.01}{\eV}. 

\begin{table}[htp]
\caption{\footnotesize  TBE/{\AVTZ} (in \si{\eV}) established for the linear systems encompassing four non-hydrogen atoms.$^a$ }
\label{Table-2}
\footnotesize
\begin{tabular}{llcccc}
ine
				&			&			&\multicolumn{2}{c}{TBEs}	&\\
Molecule			& Transition	& Nature			& New 	& Prev.			& Diff. \\
ine			
Cyanoacetylene	&$^1\Sigma^-$	($\ppi$)		&Val 	&5.79	&5.80			&-0.01\\	
				&$^1\Delta$	($\ppi$)		&Val 	&6.07	&6.07			&0.00\\	
Cyanogen			&$^1\Sigma_u^-$ ($\ppi$)		&Val 	&6.38	&6.39			&-0.01\\	
				&$^1\Delta_u$	($\ppi$)		&Val 	&6.65	&6.66			&-0.01\\	
				&$^1\Sigma_u^-$ ($\ppi$)[F]$^b$&Val&5.04	&5.05			&-0.01\\	
Diacetylene		&$^1\Sigma_u^-$ ($\ppi$)		&Val 	&5.33	&5.33			&0.00\\	
				&$^1\Delta_u$	($\ppi$)		&Val 	&5.61	&5.61			&0.00\\	
ine
\end{tabular}
\begin{flushleft}
\begin{footnotesize}
$^a${``Val'' stands for valence ESs. TBEs obtained with Eq.~\eqref{eq8} using data from the SI. The previous TBEs taken from Ref.~\citenum{Loo20a} are given for comparison.}
$^b${Fluorescence from the optimized excited-state geometry.}
\end{footnotesize}
\end{flushleft}
\end{table}

\subsection{Four non-hydrogen compounds}

For most non-linear compounds encompassing four non-hydrogen nuclei, the TBEs of the QUEST database were obtained with CCSDT/{\AVTZ} corrected using CCSDTQ/{\Pop} to take into 
account the quadruples contribution. \cite{Loo20a,Ver21} Given the results gathered in Table \ref{Table-1}, it is not clear that applying CC4/{\AVDZ} corrections would improve significantly these original 
estimates. However, for a few compounds, the QUEST TBEs do not include corrections from the quadruples as they were computed at the CCSDT/{\AVTZ} level ``only''. Hence, we have considered these 
derivatives and employed Eq.~\eqref{eq2} to determine new TBEs. The results are listed in Table \ref{Table-3} and compared to the previous estimates.

\begin{table}[htp]
\caption{\footnotesize  TBE/{\AVTZ} (in \si{\eV}) established for non-linear systems containing four non-hydrogen atoms obtained using Eq.~\eqref{eq2}.$^a$   }
\label{Table-3}
\footnotesize
\begin{tabular}{llcccc}
ine
				&			&			&\multicolumn{2}{c}{TBEs}	&\\
Molecule			& Transition	& Nature			& New 	& Prev.			& Diff. \\
ine			
Acrolein			&$^1A''$	($\npi$)	&Val &3.72	&3.78$^b$		&-0.06	\\
				&$^1A'$	($\ppi$)	&Val &6.67	&6.69			&-0.02	\\%
				&$^1A''$	($\npi$)	&Val &6.69	&6.72$^c$		&-0.03	\\%
				&$^1A'$	($\nts$)	&Ryd &7.11	&7.08			&+0.03	\\%
				&$^1A'$	($\ppi$)&Val$^d$&7.93	&7.87$^b$		&+0.06	\\
Carbonylfluoride	&$^1A_2$ ($\npi$)	&Val &7.29	&7.31$^e$		&-0.02	\\%
Cyanoformaldehyde	&$^1A''$	($\npi$)	&Val &3.82	&3.81			&+0.01	\\%
				&$^1A''$	($\ppi$)	&Val &6.43	&6.46			&-0.03	\\%
Isobutene			&$^1B_1$ ($\pts$)	&Ryd &6.48	&6.46			&+0.02	\\%
				&$^1A_1$ ($\ptp$)	&Ryd &7.01	&7.01			&0.00	\\%
Propynal			&$^1A''$	($\npi$)	&Val &3.81	&3.80			&+0.01	\\%
				&$^1A''$	($\ppi$)	&Val &5.51	&5.54			&-0.03	\\%
Thioacrolein		&$^1A''$	($\npi$)	&Val &2.10	&2.11			&-0.01	\\%
Thiopropynal		&$^1A''$ 	($\npi$)	&Val &2.02	&2.03			&-0.01	\\%
ine
\end{tabular}
\begin{flushleft}
\begin{footnotesize}
$^a${ ``Val'' and ``Ryd'' stand for valence and Rydberg ESs. Raw data are given in the SI.  The previous  TBE/{\AVTZ} values taken from Refs.~\citenum{Loo19c}, \citenum{Loo20a}, \citenum{Loo20d}, and \citenum{Ver21} 
are given for comparison.}
$^b${TBEs obtained from a basis set corrected CIPSI/{\Pop} estimate. The CCSDT/{\AVTZ} values for these two specific states are \SI{3.73}{\eV} and  \SI{8.01}{\eV}.}
$^c${Considered as \emph{unsafe} in Ref.~\citenum{Loo20a}.}
$^d${ES with a significant double excitation character.}
$^e${TBE obtained from a basis set corrected CIPSI/{\Pop} estimate.}
\end{footnotesize}
\end{flushleft}
\end{table}

Although the differences listed in the rightmost column of  Table \ref{Table-3} have different signs, the impact of CC4 is typically a small decrease of the previous estimates, hinting that CCSDT tends to 
overestimate the VTEs. This is consistent with the \SI{+0.003}{\eV} MSE reported in Table \ref{Table-1} for CCSDT as compared to CC4. For the 13 states listed in Table \ref{Table-3}, the average correction is \SI{-0.011}{\eV}
and the average absolute change is \SI{0.022}{\eV} only. More importantly, for the vast majority of cases, the changes are not larger than \SI{\pm0.03}{\eV} which was the estimated error bar in Ref.~\citenum{Loo20a}.

Nevertheless, a molecule worth discussing is acrolein. For its lowest hallmark $\npi$ transition, the CCSD, CC3, and CCSDT VTEs obtained with {\AVTZ} are \SI{3.913}{}, \SI{3.743}{}, and 
\SI{3.725}{\eV}. The new TBE, \SI{3.72}{\eV}, is  consistent with this trend, but is significantly smaller than the original CIPSI-based estimate of \SI{3.78}{\eV}. It is likely that the CIPSI extrapolation 
error bar was underestimated previously, and we believe that our current TBE is more accurate. Indeed, CC4/{\Pop} and CCSDTQ/{\Pop} values are very consistent (see Table S7). The second ES, 
a strongly dipole-allowed $^1A'$ ($\ppi$) transition, was slightly blueshifted in our original work, but the error, \SI{-0.02}{\eV}, remains low, a statement also holding for the higher-lying Rydberg 
ES of the same symmetry. The original estimate for the second $^1A''$ ($\npi$)  transition, \SI{6.72}{\eV}, was labeled as \emph{unsafe} in Ref.~\citenum{Loo20a} due to its significant 
double excitation character. Indeed, its percentage of single excitations ($\%T_1$) is only 79.4\% according to CC3/{\AVTZ}. Nevertheless, the CC4 correction obtained with {\AVDZ} is not
very large and one can likely claim that the new TBE of \SI{6.69}{\eV} stands as the most trustworthy estimate published to date for this particular ES.  The higher-lying $^1A'$ ($\ppi$) transition
of acrolein has a nature similar to the famous $A_g$ state of butadiene with a $\%T_1$ of 75\%. At the CIPSI/{\Pop} level, a value of $8.00\pm0.03$ eV was obtained previously, \cite{Loo19c}
and our CC4 value obtained with the same basis set is consistent with this result: 8.035 eV (see Table S7 in the SI). Based on CC4/{\AVDZ}, \ie,  using Eq.~\eqref{eq2}  the new TBE is 7.93
eV, upshifted by 0.06 eV as compared to the original one. Given that for the similar transition in butadiene, the changes between CC4 and CCSDTQ are trifling (Table S7), we consider
this new TBE as more trustworthy than the original one, though giving a reliable error bar is not straightforward. Finally, one 
can also note that, for carbonylfluoride, the current TBE is \SI{0.02}{\eV} smaller than the original CIPSI-based one, for which the estimated extrapolation error bar was \SI{0.02}{\eV}. \cite{Loo20a} 

\subsection{Larger systems}

\begin{table}[htp]
\caption{\footnotesize  TBE/{\AVTZ} (in eV) established for systems containing five non-hydrogen atoms.$^a$}
\label{Table-4}
\footnotesize
\begin{tabular}{llcccc}
ine
				&			&			&\multicolumn{2}{c}{TBEs}	&\\
Molecule			& Transition	& Nature			& New 	& Prev.			& Diff. \\
ine															
Cyclopentadiene	&$^1B_2$		($\ppi$)		&Val&5.54		&5.56$^b$		&-0.02	\\
				&$^1A_2$		($\pts$)		&Ryd&5.78	&5.78			&0.00	\\
				&$^1B_1$		($\ptp$)		&Ryd&6.41	&6.41			&0.00	\\
				&$^1A_2$		($\ptp$)		&Ryd&6.45	&6.46			&-0.01	\\
				&$^1B_2$		($\ptp$)		&Ryd&6.56	&6.56			&0.00	\\	
				&$^1A_1$	 	($\ppi$)		&Val&6.45		&6.52$^c$		&-0.07	\\
Difluorodiazirine	&$^1B_1$ 	($\npi$)		&Val&3.73		&3.74			&-0.01	\\
				&$^1A_2$ 	($\ppi$)		&Val&6.99		&7.00			&-0.01	\\
				&$^1B_2$		(n.d.)$^d$		&Ryd&8.50	&8.52			&-0.02	\\
Furan			&$^1A_2$		($\pts$)		&Ryd&6.10	&6.09			&+0.01	\\
				&$^1B_2$		($\ppi$)		&Val&6.35		&6.37			&-0.02	\\
				&$^1A_1$		($\ppi$)		&Val&6.53 	&6.56			&-0.03	\\
				&$^1B_1$		($\ptp$)		&Ryd&6.65	&6.64			&+0.01	\\
				&$^1A_2$		($\ptp$)		&Ryd&6.82	&6.81			&+0.01	\\
				&$^1B_2$		($\ptp$)		&Ryd&7.25	&7.24			&+0.01	\\
Pyrrole			&$^1A_2$ 	($\pts$)		&Ryd&5.23	&5.24			&-0.01	\\
				&$^1B_1$ 	(n.d.)			&Ryd&5.97	&6.00			&-0.03	\\
				&$^1A_2$ 	($\ptp$)		&Ryd&6.01	&6.00			&+0.01	\\
				&$^1B_1$		(n.d.)$^e$		&Ryd&6.09	&				&		\\
				&$^1B_2$ 	($\ppi$)		&Val&6.24		&6.26			&-0.02	\\
				&$^1A_1$ 	($\ppi$)		&Val&6.27		&6.30			&-0.03	\\
				&$^1B_2$ 	($\ptp$)		&Ryd&6.82	&6.83			&-0.01	\\
Thiopehene		&$^1A_1$ 	($\ppi$)		&Val&5.62		&5.64			&-0.02	\\
				&$^1B_2$ 	($\ppi$)		&Val&5.94		&5.98			&-0.04	\\
				&$^1A_2$ 	($\pts$)		&Ryd&6.13	&6.14			&-0.01	\\
				&$^1B_1$ 	($\ptp$)		&Ryd&6.12	&6.14			&-0.02	\\
				&$^1A_2$ 	($\ptp$)		&Ryd&6.23	&6.21			&+0.02	\\
				&$^1B_1$ 	($\pts$)		&Ryd&6.49	&6.49			&0.00	\\
				&$^1B_2$ 	($\ptp$)$^f$	&Ryd&7.27	&7.29			&-0.02	\\
				&$^1A_1$		($\ppi$)$^f$	&Val&7.30		&7.31$^c$		&-0.01	\\
ine
\end{tabular}
\begin{flushleft}
\begin{footnotesize}
$^a${ ``Val'' and ``Ryd'' stand for valence and Rydberg ESs. TBEs established with Eq.~\eqref{eq3}, except for difluorodiazirine and furan for which  Eq.~\eqref{eq2} is used.
Raw data are given in Table S10 in the SI. The previous TBEs taken from Refs.~\citenum{Loo20a} and \citenum{Loo20d} are also listed.}
$^b${The most recent TBE, based on a basis set corrected CIPSI/{\Pop} value, is \SI{5.54}{\eV}, see Ref.~\citenum{Ver21}.}
$^c${Considered as \emph{unsafe} in Ref.~\citenum{Loo20a}.}
$^d${Incorrectly labeled as valence in Ref.~\citenum{Loo20d}.}
$^e${Not considered previously.}
$^f${Non-negligible valence/Rydberg mixing.}
\end{footnotesize}
\end{flushleft}
\end{table}

For the largest systems of Figure \ref{Fig-1}, CC4 has the advantage to provide a systematic path towards high-accuracy for systems that are beyond reach at the CCSDTQ level.
Let us first illustrate this with a set of 30 ESs extracted from systems with five (non-hydrogen) atom (Table \ref{Table-4}). To the best of our knowledge, for all five molecules, our 
results provide the first VTEs including quadruples corrections. As one can see, the differences with the previous TBEs are small, and are typically slightly negative, confirming the outcome 
of the previous section showing that CCSDT/{\AVTZ} yields slightly too large VTEs.  The average absolute correction for the data of Table \ref{Table-4} is nonetheless very small, \SI{0.02}{\eV}, 
which confirms that quality of the original TBEs from the QUEST database. 

Let us now take a look at the most problematic cases. Unsurprisingly, the $^1A_1$ ($\ppi$) transition of cyclopentadiene is significantly redshifted by the CC4 correction (\SI{-0.07}{\eV}). This transition 
was previously classified as \emph{unsafe}, \cite{Loo20a} due to its $\%T_1$ value of 78.9\% and its similar nature to the famous $^1A_g$ ES of butadiene. For the latter compound, the difference 
between the CIPSI/{\Pop} and CCSDT/{\Pop} estimates is \SI{-0.08}{\eV}, \cite{Loo19c} which suggests that the CC4 correction of \SI{-0.07}{\eV} for the $^1A_1$ ES of cyclopentadiene is very reasonable.  
The lowest $B_2$ ES of thiophene is subject to a \SI{-0.04}{\eV} shift despite having also a very large single excitation character ($\%T_1 = 91.5\%$).  All other ESs are even less affected by the CC4 corrections
with variations of $\pm$\SI{0.03}{\eV} at most.

We now consider three highly-symmetric six-membered rings, namely benzene, pyrazine, and tetrazine, for which CC4/{\Pop} calculations are still doable. The results obtained for a large number
of ESs are listed in Table \ref{Table-5}. For these three compounds, this Table is, as far as we are aware of, the first to propose CC-based VTEs including corrections from iterative quadruples. 

\begin{table}[htp]
\caption{\footnotesize  TBE/{\AVTZ} (in eV) established for benzene, pyrazine, and tetrazine.$^a$}
\label{Table-5}
\footnotesize
\begin{tabular}{llcccc}
ine
				&			&			&\multicolumn{2}{c}{TBEs}	&\\
Molecule			& Transition	& Nature			& New 	& Prev.			& Diff. \\
ine															
Benzene			&$^1B_{2u}$	($\ppi$)				&Val &5.05	&5.06			&-0.01\\
				&$^1B_{1u}$	($\ppi$)				&Val &6.43	&6.45			&-0.02\\
				&$^1E_{1g}$	($\pts$)				&Ryd &6.52	&6.52			&0.00\\
				&$^1A_{2u}$	($\ptp$)				&Ryd &7.08	&7.08			&0.00\\
				&$^1E_{2u}$	($\ptp$)				&Ryd &7.15	&7.15			&0.00\\
				&$^1A_{1u}$	($\ptp$)$^b$			&Ryd &7.23	&				&\\
				&$^1E_{1u}$	($\ppi$)$^b$			&Val &7.17	&				&\\
Pyrazine			&$^1B_{3u}$	($\npi$)				&Val &4.14	&4.15			&-0.01\\
				&$^1A_{u}$	($\npi$)				&Val &4.97	&4.98			&-0.01\\
				&$^1B_{2u}$	($\ppi$)				&Val &4.99	&5.02			&-0.03\\
				&$^1B_{2g}$	($\npi$)				&Val &5.68	&5.71			&-0.03\\	
				&$^1A_{g}$	($\nts$)				&Ryd &6.66	&6.65			&+0.01\\	
				&$^1B_{1g}$	($\npi$)				&Val &6.70	&6.74			&-0.04\\	
				&$^1B_{1u}$	($\ppi$)				&Val &6.85	&6.88			&-0.03\\	
				&$^1B_{1g}$	($\pts$)				&Ryd &7.20	&7.21			&-0.01\\
				&$^1B_{2u}$	($\ntp$)				&Ryd &7.27	&7.24			&+0.03\\
				&$^1B_{1u}$	($\ntp$)				&Ryd &7.45	&7.44			&+0.01\\	
				&$^1B_{1u}$	($\ppi$)				&Val &7.94	&7.98$^c$		&-0.04\\
Tetrazine			&$^1B_{3u}$	($\npi$)				&Val &2.46	&2.47			&-0.01\\
				&$^1A_{u}$	($\npi$)				&Val &3.68	&3.69			&-0.01\\
				&$^1A_{g}$	($n,n\ra\pis,\pis$)		&Val &		&4.61$^d$		&\\	
				&$^1B_{1g}$	($\npi$)				&Val &4.87	&4.93			&-0.06\\
				&$^1B_{2u}$	($\ppi$)				&Val &5.17	&5.21			&-0.04\\
				&$^1B_{2g}$	($\npi$)				&Val &5.50	&5.45			&-0.05\\
				&$^1A_{u}$	($\npi$)				&Val &5.51	&5.53			&-0.02\\
				&$^1B_{3g}$	($n,n\ra\pis,\pis$)		&Val &		&6.15$^d$		&\\	
				&$^1B_{2g}$	($\npi$)				&Val &6.05	&6.12			&-0.07\\
				&$^1B_{3g}$	($\nts$)$^b$			&Ryd &6.47	&				&\\
				&$^1B_{3u}$	($\ppi$)$^b$			&Val &6.67	&				&\\
				&$^1B_{1g}$	($\npi$)				&Val &6.89	&6.91			&-0.02\\
ine
\end{tabular}
\begin{flushleft}
\begin{footnotesize}
$^a${``Val'' and ``Ryd'' stand for valence and Rydberg ESs. TBEs established using Eq.~\eqref{eq3}. Raw data are given in Table S11 in the SI. The previous TBEs taken 
from Ref.~\citenum{Loo20a}  are also listed.}
$^b${Not considered previously.}
$^c${Considered as \emph{unsafe} in Ref.~\citenum{Loo20a}.}
$^d${Genuine doubly-excited states with TBEs obtained at the NEVPT2/{\AVTZ} level and considered as \emph{unsafe} in Ref.~\citenum{Loo20a}.}
\end{footnotesize}
\end{flushleft}
\end{table}

For benzene, we consider three valence and four Rydberg ESs. These ESs have a very strong single excitation character with $\%T_1 > 92\%$, except for the lowest 
$^1B_{2u}$ transition ($\%T_1 = 86\%$). In all cases, one finds, as most of the five-membered cycles, that CC4/{\Pop} provides small negative or null corrections, with a maximal amplitude 
of \SI{-0.02}{\eV}, even for the lowest-energy transition.

Eleven ESs of pyrazine are listed in Table \ref{Table-5}, including four Rydberg states. The corrections to previous CCSDT-based TBEs are either positive, null or negative, but most are within the expected
error bar of the original estimates. Nevertheless, for the lowest $1^1B_{1g}$ ES, corresponding to a $\npi$ excitation, a relatively large correction of \SI{-0.04}{\eV} is noticed. This
transition has a rather low $\%T_1$ value of $84\%$, which however does not translate into a large CC3/CCSDT difference (ca.~\SI{0.01}{\eV} for the three basis sets, see the SI). Likewise
the highest-lying $1^1B_{1u}$ ES also undergoes a  \SI{-0.04}{\eV} correction, but this state was rated \emph{unsafe} previously, \cite{Loo20a,Ver21} due to an unusually
large CC3/CCSDT difference.  Four other ESs have their VTEs corrected by $\pm$\SI{0.03}{\eV} by CC4, a shift corresponding to the expected error bar of the original TBEs.

In tetrazine, we consider 12 ESs, including two with a pure double excitation character, for which NEVPT2 was employed in the QUEST database to obtain TBEs, 
whereas,  for the other transitions, CCSDT/{\AVTZ} or basis set corrected CCSDT/{\AVDZ} values were originally selected as TBEs. \cite{Loo20a} All the latter were rated as \emph{safe} 
as the energy differences between CCSDT and CC3 were smaller than \SI{0.03}{\eV} despite $\%T_1$ values often smaller than 90\%. \cite{Loo20a,Ver21} For several ESs, small negative 
CC4 corrections, \SI{-0.01}{} or \SI{-0.02}{\eV}, are obtained.  For the lowest $\ppi$ transition, the correction is only slightly larger, i.e., \SI{-0.04}{\eV}. However, for three $\npi$ 
transitions ($1^1B_{1g}$, $1^1B_{2g}$, and $2^1B_{2g}$), much larger changes (\SI{-0.06}{}, \SI{-0.05}{}, and \SI{-0.07}{\eV}) are induced by the inclusion of quadruples, indicating that our 
original assessment of the TBE quality was optimistic.  Interestingly, for these three transitions, one has $\%T_1 < 85\%$, in contrast with the other singly-excited ESs for which  $\%T_1 > 85\%$. 
This observation suggests that the $85\%$ barrier might be the limit for which CCSDT can be considered trustworthy, irrespective of the difference between CC3 
and CCSDT. For the two doubly-excited states of tetrazine listed in Tables \ref{Table-5} and S11, that are both  characterized by $\%T_1 < 1\%$, \cite{Loo20a} the differences between CC4/{\Pop} and CCSDT/{\Pop} 
are very significant, that is, \SI{-0.79}{\eV} ($A_g$) and \SI{-0.99}{\eV} ($B_{3g}$). These values are nonetheless typical for these challenging ESs with a dominant contribution from the doubly-excited 
determinants. Applying Eq.~\eqref{eq3} delivers VTEs of \SI{5.17}{} and \SI{6.44}{\eV}. These two results logically remain higher in energy than the corresponding NEVPT2/{\AVTZ} values of \SI{4.61}{} 
and \SI{6.15}{\eV}, that we consider as the most accurate  TBEs available to date, though with an error bar of approximately \SI{\pm0.1}{\eV}.

Finally, let us briefly discuss the challenging case of hexatriene (Table \ref{Table-6}). The original TBEs for this small polyene have been obtained via Eq.~\eqref{eq6}: 5.37, 5.62, 5.79, and 5.94 eV for the 
lowest $^1B_{u}$, $^1A_{g}$, $^1A_{u}$, and $^1B_{g}$ ESs, respectively. \cite{Ver21} All were considered \emph{safe}, except the second one that has a very significant contribution from the double excitations with 
$\%T_1 = 65\%$ (i.e., $10\%$ less than in butadiene). Using Eq.~\eqref{eq5} to include CC4 corrections leads to improved TBEs of 5.34, 5.46, 5.79, and 5.93 eV. Only the $^1A_{g}$
VTE undergoes a significant downshift  (\SI{-0.16}{\eV}) due to the quadruples. Logically, this  \SI{5.46}{\eV} value can be considered as an upper bound and we roughly approximate the exact value 
to be of the order of \SI{5.43}{\eV}. To support this crude estimate, we looked at the equivalent transition in the smaller polyene, butadiene. Indeed, in butadiene, the difference between the CC4/{\Pop} and FCI/{\Pop} values is \SI{-0.01}{\eV}; 
the  correction should be larger for hexatriene given its smaller $\%T_1$ value. In addition, as can be seen in Ref.~\citenum{Loo19c}, the NEVPT2/{\Pop} value is \SI{0.20}{\eV} larger than the FCI/{\Pop} result for
butadiene, and applying such rigid shift to NEVPT2/{\AVTZ} VTE of hexatriene, \cite{Loo19c} would yield an estimate of \SI{5.44}{\eV}. Although it would not be suited to rate this  new TBE/{\AVTZ} of \SI{5.43}{\eV}
as \emph{safe}, it is likely one of the most accurate estimate published to date for this dark transition.

\begin{table}[htp]
\caption{\footnotesize  TBE/{\AVTZ} (in eV) established for hexatriene.$^a$}
\label{Table-6}
\footnotesize
\begin{tabular}{llcccc}
ine
				&			&			&\multicolumn{2}{c}{TBEs}	&\\
Molecule			& Transition	& Nature			& New 		& Prev.		& Diff. \\
ine															
Hexatriene		&$^1B_{u}$	($\ppi$)	&Val		&5.34		&5.37		&-0.03\\
				&$^1A_{g}$	($\ppi$)	&Val		&5.46/5.43$^c$	&5.62$^b$	&-0.16/-0.19\\
				&$^1A_{u}$	($\pts$)	&Ryd	&5.79		&5.79		&0.00\\
				&$^1B_{g}$	($\ptp$)	&Ryd	&5.93		&5.94		&-0.01\\
ine
\end{tabular}
\begin{flushleft}
\begin{footnotesize}
$^a${``Val'' and ``Ryd'' stand for valence and Rydberg ESs. TBEs established using Eq.~\eqref{eq5}. Raw data are given in Table S11 in the SI. The previous TBEs taken 
from Ref.~\citenum{Ver21} are also listed.}
$^b${Considered as \emph{unsafe} in Ref.~\citenum{Ver21}.}
$^c${See text.}
\end{footnotesize}
\end{flushleft}
\end{table}

\section{Conclusions and outlook}

Three main problems were tackled in the present study. Our first aim was to confirm that CC4 provides very accurate VTEs. To this end, we considered 43 molecules containing between two 
and four (non-hydrogen) atoms and computed their ESs with three Gaussian basis sets containing diffuse functions, i.e., {\Pop}, {\AVDZ}, and {\AVTZ}. We started by defining more than 200 
CCSDTQ reference values. Excluding the few pathological pure doubly-excited states of $(n,n)\ra(\pis,\pis)$ nature but conserving all other transitions that have a non-negligible double excitation 
character ($2^1A_g$ in butadiene, $2^1A"$ in acrolein, etc), we showed that CC4 is an excellent approximation of CCSDTQ with a mean absolute error of \SI{0.003}{\eV}. This is a totally negligible 
deviation for the vast majority of chemical studies ($< 0.1$ kcal.mol$^{-1}$). We also showed that neither CC3 nor CCSDT could deliver the same level of accuracy for the very same set of ESs.

Second, we investigated the performance of additive basis set correction schemes in the EOM-CC context using the same set of ESs as in the first part. It appeared that these popular
correction strategies indeed statistically improve the quality of the VTEs. More interestingly, such basis set corrections were found particularly powerful when starting from CC3/{\AVTZ} 
or CCSDT/{\AVTZ}, as they allowed estimating the true  CC4/{\AVTZ} or CCSDTQ/{\AVTZ} VTEs with average errors of approximately \SI{0.005}{\eV}. Of course, it is essential to confirm these trends 
for larger compounds, but the cost of CC4/{\AVTZ} and/or CCSDTQ/{\AVTZ} unfortunately prevents us of doing it as of today.

Finally, CC4 was employed to improve previous CCSDT-based TBEs of the QUEST database. \cite{Ver21} For benzene, cyclopentadiene, difluorodiazirine, furan, hexatriene, pyrazine, pyrrole, tetrazine, 
and thiophene, it was possible to perform CC4/{\Pop} and/or CC4/{\AVDZ} calculations whereas CCSDTQ remains beyond reach. The corrections obtained were, in most cases, 
smaller than the expected average error of the original TBEs (ca.~\SI{\pm0.03}{\eV}). Nevertheless, most corrections were null or negative, hinting at a slight overestimation trend in our original
TBEs. In addition, some of the estimates originally viewed as \textit{unsafe} can be now considered as trustworthy ($2^1A"$ in acrolein, $2^1A_1$ in cyclopentadiene, $3^1A_1$ in thiophene, etc) 
whereas a few optimistic assessments have been revised, resulting in more accurate TBEs (in particular for the valence $B_{1g}$ and $B_{2g}$ ESs of tetrazine). 

We are currently pursuing our efforts to improve the quality, size, and diversity of the QUEST database, so as to provide the most trustworthy reference values possible to the community. We plan
to publish, within a year or two, an expanded and improved version of the QUEST database.

\section*{Acknowledgements}
The authors would like to warmly thank Martial Boggio-Pasqua, Anthony Scemama, and Claudia Filippi for numerous fruitful discussions. PFL thanks the European Research 
Council (ERC) under the European Union's Horizon 2020 research and innovation programme (grant agreement no.~863481) for financial support. DJ is indebted to the CCIPL 
computational center installed in Nantes for (the always very) generous allocation of computational time. 
\section*{Supporting Information Available}
Raw VTEs. Geometries.

\bibliography{biblio-new}

\end{document}